\documentclass{pnasapp}

\usepackage{pnastwof}
\usepackage{color,graphicx}
\usepackage{array} 
\usepackage{url}
\usepackage{xcolor}
\usepackage{catchfilebetweentags}
\usepackage{amssymb,amsfonts,amsmath}

\usepackage[comma,sort&compress]{natbib}
\def \aap{A\&A}

\def \apjl{ApJ}

\def \apj{ApJ}

\def \araa{ARA\&A}

\def \jqsrt{J.~Quant.~Spectrosc.~Radiat.~Transfer}

\def \nat{Nat}

\newcommand{\dep}[2]{\ensuremath{\left(#1\right)^{#2}}}

\definecolor{lightblue}{rgb}{0.80,0.85,0.92}
\definecolor{darkblue}{rgb}{0.00,0.41,0.66}

\newcommand\blfootnote[1]{%
  \begingroup
    \renewcommand\thefootnote{}\footnotetext{#1}%
      \addtocounter{footnote}{-1}%
        \endgroup
      \addtocounter{footnote}{1}%
}

\begin{document}

\title{Some inconvenient truths about `biosignatures' involving two chemical species on Earth-like exoplanets}

\author{Hanno Rein\affil{1}{University of Toronto, Department of Environmental and Physical Sciences, Toronto, Ontario M1C 1A4}$^{\rm{a}}$, 
Yuka Fujii\affil{2}{Earth-Life Science Institute, Tokyo Institute of Technology,  Ookayama, Meguro, Tokyo 152-8550, Japan}
\and David S. Spiegel\affil{3}{Astrophysics Department, Institute for Advanced Study, Princeton, NJ  08540}}
\footlineauthor{Rein, Fujii, and Spiegel}

\contributor{Edited by Neta A. Bahcall, Princeton University, Princeton, NJ, and approved March 27, 2014 (received for review February 1, 2014)}
\maketitle

\begin{article}

\begin{abstract}
The detection of strong thermochemical disequilibrium in the atmosphere of an extrasolar planet is thought to be a potential biosignature.
In this article we present a new kind of false positive that can mimic a disequilibrium or any other biosignature that involves two chemical species. 
We consider a scenario where the exoplanet hosts a moon that has its own atmosphere and neither of the atmospheres is in chemical disequilibrium.
Our results show that the integrated spectrum of the planet and the moon closely resembles that of a single object in strong chemical disequilibrium.
We derive a firm limit on the maximum spectral resolution that can be obtained for both directly-imaged and transiting planets.
The spectral resolution of even idealized space-based spectrographs that might be achievable in the next several decades is in general insufficient to break the degeneracy.
Both chemical species can only be definitively confirmed in the same object if absorption features of both chemicals can be unambiguously identified and their combined depth exceeds~100\%.
\end{abstract}

\keywords{astrobiology | detection of life | biomarker | exomoon | habitability }
\blfootnote{$^{\rm{a}}$To whom correspondence should be addressed. E-mail: \url{hanno.rein@utoronto.ca}.} 

\noindent\fcolorbox{darkblue}{lightblue}{\parbox{\dimexpr \linewidth-2\fboxsep-2\fboxrule}{%
\abstractfont \color{darkblue} { Significance\\[5pt]} 
The search for life on planets outside our own Solar System is among the most compelling quests that humanity has ever undertaken. 
An often suggested method of searching for signs of life on such planets involves looking for spectral signatures of strong chemical disequilibrium.
This article introduces an important potential source of confusion associated with this method.
Any exoplanet can host a moon that contaminates the planetary spectrum.
In general, we will be unable to exclude the existence of a moon.
By calculating the most optimistic spectral resolution in principle obtainable for Earth-like planets, we show that inferring a biosphere on an exoplanet might be beyond our reach in the foreseeable future.
}}

\section{Introduction}
With almost a thousand confirmed exoplanets \cite[Open Exoplanet Catalogue,][]{Rein2012}, the prospects of detecting signs of a biosphere on a body outside our own solar system are more promising than ever before.
However, there are still huge technological and theoretical challenges to overcome before one can hope to make a clear detection of life on an exoplanet.
In this article, we discuss one of these complications, the possibility of false positives due to the presence of an exomoon orbiting the exoplanet.

There are many ways that life on an exoplanet might affect the planet's appearance, ranging from deliberate signals from intelligent civilizations \cite{tarter2001} to subtler signs of simple life.
In order to characterize an extrasolar world as fully as possible, we ideally would measure its spectrum as a function of time in both the optical and the infrared parts of the spectrum \cite[e.g.][]{cowan+agol2009,kawahara+fujii2011,fujii+kawahara2012,fujii_et_al2013}.
For example, spectral evidence of water could suggest that a planet might be habitable.
It has also been suggested that an intriguing indication of life might be an increase in the planet's albedo toward the infrared part of the spectrum, which on Earth can be associated with vegetation \cite{seager_et_al2005}.
However, these features alone would not be smoking-gun proof of the presence of life.
The terms `biomarker' and `biosignature' generally refer to chemicals or combinations of chemicals that {\it could} be produced by life and that {\it could not} be (or are unlikely to be) produced abiotically; hereafter, we use these terms interchangeably.
If biosignature gases are detected in the spectrum of an exoplanet, the probability that they actually indicate life depends both on the prior probability of life \cite{spiegel+turner2012} and on the probability that the observed spectroscopic feature could be produced abiotically.
The latter possibility is the subject of this paper.

Byproducts of metabolism are often thought of as the most promising biomarker \cite{Kaltenegger2002,Kaltenegger2007,Kaltenegger2010,Seager2012,Seager2013a,Seager2013b,Kasting2013}. 
More specifically, an extreme thermodynamic disequilibrium of two molecules in the atmosphere is considered a biosignature \cite{Lederberg1965, Lovelock1965, Segura2005}.
An example of two such species is the simultaneous presence of O$_2$ and a reduced gas such as CH$_4$.
It is important to point out that a disequilibrium in a planet's atmosphere should not be considered as clear evidence for life.\footnote{Also note that the Earth might have never had a phase of strong, observable O$_2$/CH$_4$ disequilibrium \cite{Marais2002}.}
There is a long list of abiotic sources that could also create a disequilibrium such as impacts~\cite{Kasting1990}, photochemistry~\cite{Zahnle2008}, and geochemistry~\cite{Seager2013b}. 

In this article, we describe a new scenario for a possible false positive biosignature.
If the exoplanet hosts a moon that has an atmosphere itself, the simultaneous observation of the planet and moon modifies the observed spectrum  \cite[see also][]{Cabrera2007} and can produce a signal that looks like a disequilibrium in one atmosphere, but is in fact created by two atmospheres blended together.
It might be extremely difficult to discern that an exoplanet even has a moon, let alone that one component of a two-chemical biosignature comes from the moon instead of the planet.

\begin{figure*}[th!]
\begin{center}
\resizebox{\columnwidth}{!}{\includegraphics{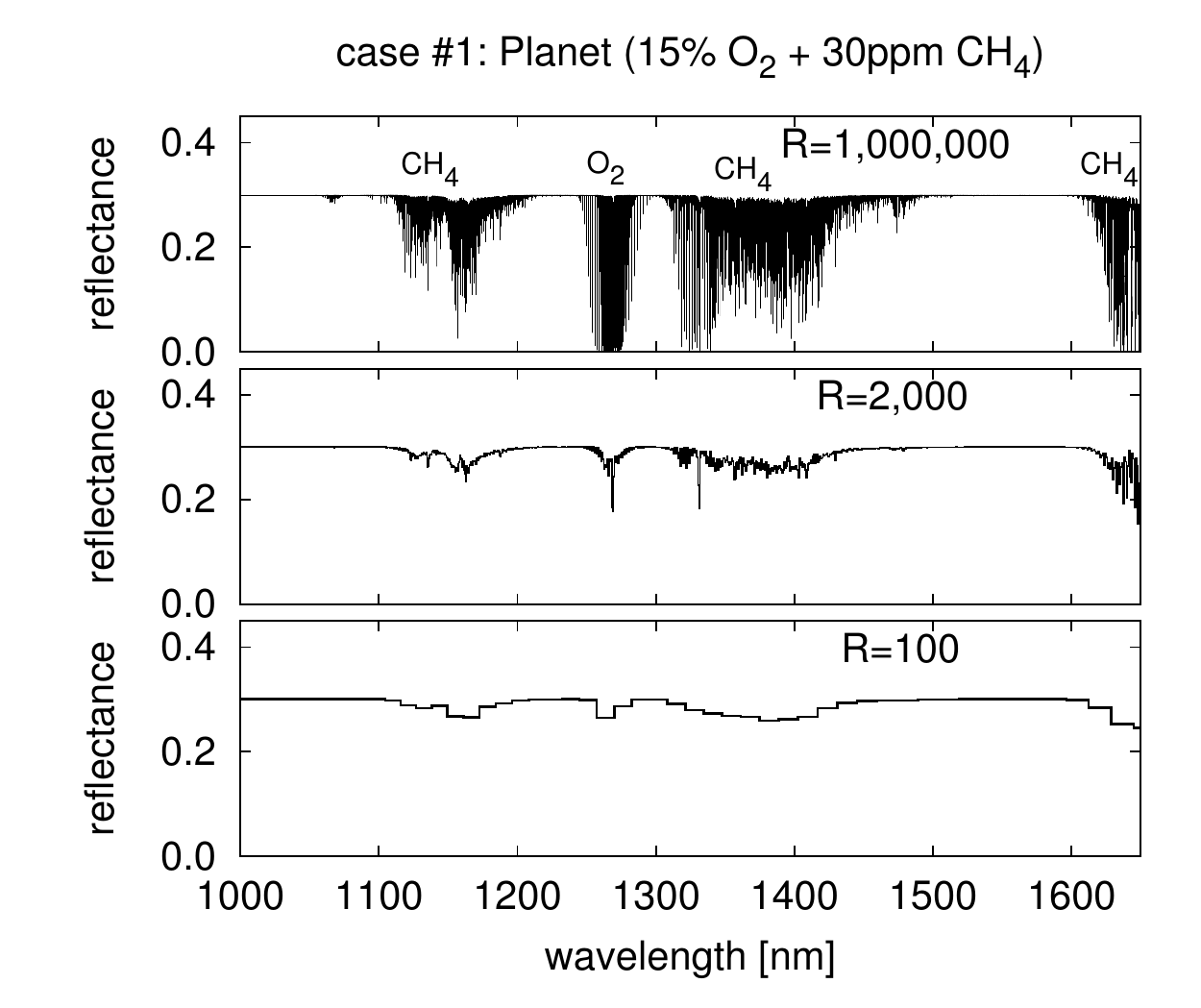}}
\resizebox{\columnwidth}{!}{\includegraphics{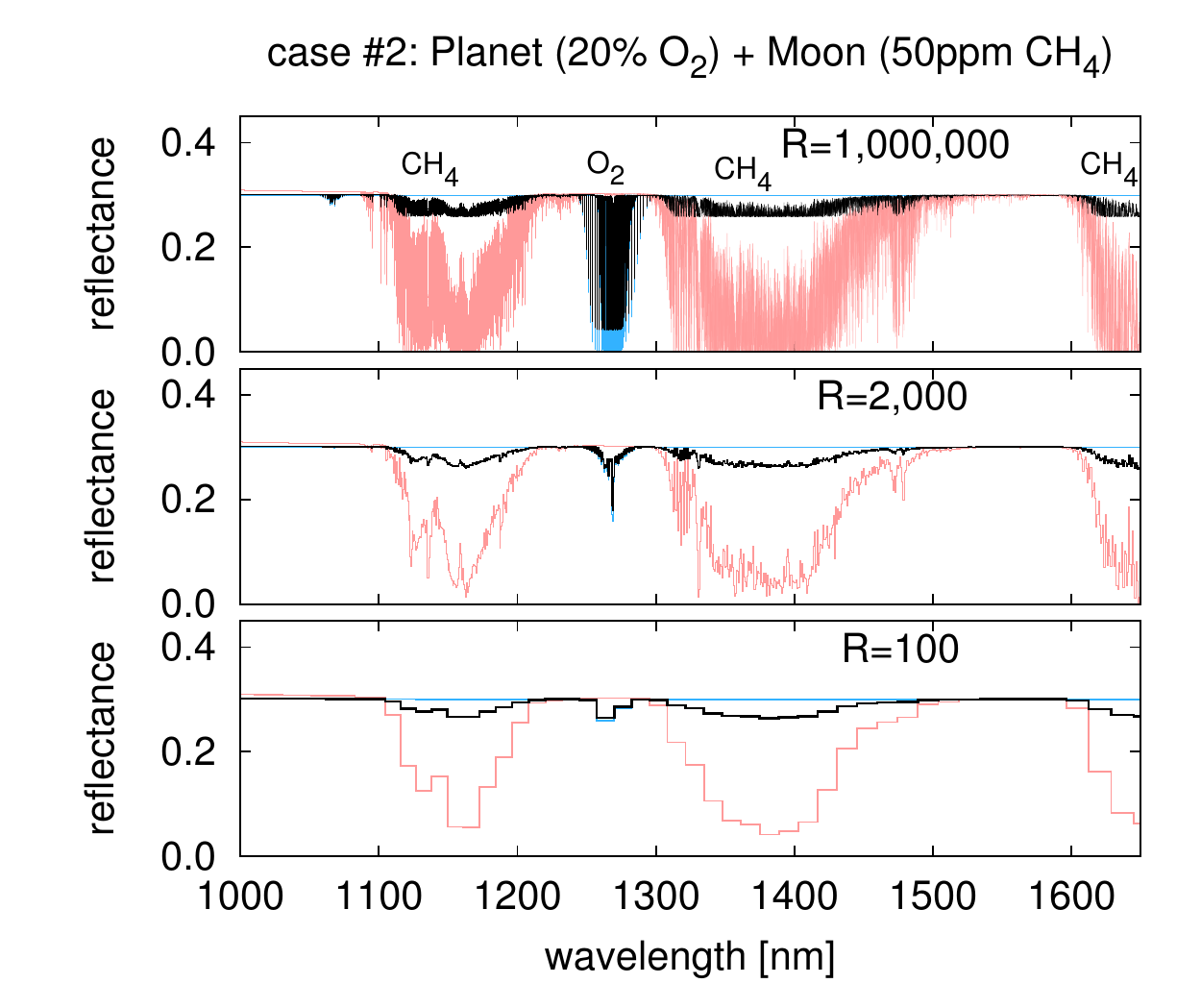}}
\end{center}
\caption{Model spectra for cases \#1 and \#2 with varying resolution.
      (Left) Model spectra of a planet with 15\% O$_2$ and 30ppm CH$_4$ (case \#1).
      (Right) Black lines: Combined spectra of a planet with 20\% O$_2$ and a moon with 50ppm CH$_4$.
      Blue lines: Model spectra of a planet with 20\% O$_2$. Red lines: Model spectra of a moon with 50ppm CH$_4$.\label{fig:spcomp}
      }
\end{figure*}

The outline of this article is as follows.
We first describe our model atmospheres and present simulated spectra.
Using those synthetic spectra, we show that the combined spectrum from an oxygen-rich atmosphere such as that of the Earth and a methane rich atmosphere such as that of Titan indeed looks like it could have come from a single atmosphere with a strong disequilibrium.
We then calculate a strong upper limit on the spectral resolution of such a system as observed from Earth under ideal conditions with a plausibly-sized space telescope.
Our estimate shows that the spectral resolution $R \equiv \lambda/d \lambda$ for such a system is unlikely to exceed~$\sim$1600 with foreseeable technology.
Given this maximum-possible resolution, discriminating between a single planet and a planet-moon system is in general unlikely to be possible.\footnote{
If an exomoon transits its planet, its ``secondary-eclipse'' (when it dips behind the planet) offers a way to break the degeneracy, because it presents an opportunity to obtain the spectrum of the planet alone.
Transits, however, are unlikely.
For instance, the Earth's Moon transits the Earth for only $\sim$2\% of randomly-oriented observers.
}
Nevertheless, we conclude with a summary and a positive outlook with two possibilities that can provide genuine biosignatures.
The first possibility is to find a single chemical species that is sufficient to indicate life.
The second one requires the unambiguous identification of both species' absorption features and the combined depth of the features needs to exceed~100\%.

\section{Models}

In order to demonstrate how a planetary spectrum with two `biosignature' molecules could be spoofed by an unseen moon, we compare two simulated spectra: (case \#1) that of a planet that has both O$_2$ and CH$_4$ in its atmosphere, and (case \#2) that of a (spatially-unresolved) planet+moon system where each body contains either O$_2$ or CH$_4$ but not both.
We calculate 1-D line-by-line radiative transfer based on \texttt{DISORT} \cite[Discrete Ordinates Radiative Transfer Program for a Multi-Layered Plane-Parallel Medium;][]{Stamnes1988} and the spectroscopic database \texttt{HITRAN} \cite{Rothman2013}. 
    
Although absorption by other chemical species is naturally expected for habitable planet candidates, we exclude all species besides O$_2$ and CH$_4$ from our model, because doing this clarifies our argument.  
Nevertheless, water, for instance, has strong absorption absorption features that overlap with some oxygen and methane bands.  
This can further complicate inferring the presence of any biosignatures in a spectrum~(we address this in the Supporting Information).

The geometric configuration is fixed at $\theta_0=\theta_1=45^{\circ }$ and $\phi =0^{\circ }$ where $\theta _0$, $\theta _1$, and $\phi $ are the zenith angle of the incident light, that of the observation, and the relative azimuthal angle\footnote{$\phi =0^{\circ}$ represents the forward scattering.}, respectively, as a representative geometry for a planet at quadrature. 
We assume that the model atmospheres are characterized by the ideal gas equation of state, hydrostatic equilibrium, and (dry) adiabatic temperature profiles from the 300-K surface up to 150~K:
\begin{equation}
P = \frac{\rho }{\mu } \mathcal{R} T, \;\;\; \frac{dP}{dz} = -\rho g, \;\;\; \frac{dT}{dz} = -\Gamma = -\frac{g}{c_p} \, , \label{eq:atmprof}
\end{equation}
where $\mathcal{R}$ is the ideal gas constant, $\rho$, and $z$ have their usual meanings, and the other symbols are described in Table 1.

Real planetary atmospheres do not follow an adiabat above the tropopause and therefore do not continue to get arbitrarily colder with altitude.  
Stratospheric temperature profiles can be complicated, but for illustrative purposes we simply take the model atmosphere to be isothermal above the altitude at which the adiabat reaches $T = 150$~K.
Furthermore, for the sake of simplicity, we neglect the effects of clouds or hazes.
The physical parameters to specify the atmospheric profiles are as listed in Tables 1 and 2, where the planet mimics a (cloud-less) Earth while the moon has Titan-like properties. 
We can neglect the Doppler shift of the spectral lines as the relative Doppler shift of the Earth-moon system in an edge-on orbit is only~$\Delta \lambda /\lambda \sim 3\times 10^{-6}$.

Figure \ref{fig:spcomp} compares the spectrum of a planet with 15\% O$_2$ and 30ppm CH$_4$ (case \#1) to that of a planet with 20\% O$_2$ plus a moon with 50 ppm CH$_4$ (case \#2).
For case \#2, we independently calculated the spectra for the planet and for the moon, find the sum of the two, and normalized by the total area of the planetary and moon disks $r_\oplus^2+(0.4r_\oplus)^2 = 1.16r_\oplus^2$, where $r_\oplus$ is the radius of the Earth. 

At very high resolution ($R=1,000,000$, top panels in Figure~\ref{fig:spcomp}) where lines are well resolved, the difference between case~\#1 and~\#2 is clear. 
In the spectrum of a planet alone, many of the line cores of both O$_2$ and CH$_4$ hit zero, while in the spectrum of a planet plus a moon, the lines are peculiarly cut well above 0. 
In the latter case, many of the O$_2$ and CH$_4$ lines are actually saturated in the spectrum of the planet and that of the moon, respectively.
After adding the two, the scattered light from the other body contributes as an offset.

At lower resolutions, however, the two spectra are almost indistinguishable. 
The saturated line cores are smoothed off, resulting in very similar shallow absorption bands. 
The direct comparison is shown in Figure \ref{fig:spcomp_R100} with the uncertainty bars corresponding to signal-to-noise ratio (SNR) of $\sim$10.
Although subtle differences exist in the shapes of the band and the slope of the continuum, it would be impossible to discern such differences from moderate-resolution ($R=100$) observations without a priori knowledge of the detailed chemistry and physical properties of the target bodies.
We therefore conclude that the presence of an unseen moon may be responsible for the apparent co-existence of two species in disequilibrium.

\begin{table}
{\fignumfont Table 1. \;\;Basic Assumptions for Planet and Moon \hfill}

\vspace{3pt}
\centering
\begin{tabular}{l|ccc} \
&Symbol & Planet	& Moon \\ \hline
radius  & r & r$_{\oplus}$ & 0.4r$_{\oplus}$ \\
surface gravity	& $g$ & 9.8 m/s$^2$	& 1.35 m/s$^2$	\\
surface temperature & $T_0$ & 300 K	& 300 K	\\
surface pressure & $P_0$ & 1013 mbar	& 1500 mbar	\\
surface albedo & a & 0.3 & 0.3\\
mean molecular weight & $\mu $ & 28.8 & 27.0  \\
heat capacity & $c_p$ & 1.0 & 1.0 \\ 
dry adiabatic lapse rate & $\Gamma $ & 9.8 K/km & 1.35 K/km \\ 
cloud or haze & --- & no & no 
\end{tabular}
\end{table}%
\begin{table}
{\fignumfont Table 2. \;\;Models to Compare\hfill}

\vspace{3pt}
\centering
\begin{tabular}{l|c|cc} 
& case \#1 & \multicolumn{2}{c}{case \#2} \\ \hline
target  & Planet & Planet & Moon  \\
composition & 15\% O$_2$+30ppm CH$_4$ & 20\% O$_2$ & 50ppm CH$_4$ \\
normalization & $r_\oplus^2$ & \multicolumn{2}{c}{$1.16r_\oplus^2$} 
\end{tabular}
\label{tab:cases}
\end{table}%

There is one other possibility to confirm the presence of two species in the same object.
Unfortunately, this requires one to be able to unambiguously identify absorption bands, which is hard or even impossible in a low resolution spectra without any a-priori knowledge of the atmosphere's composition. 
Let us ignore these difficulties for now and assume that we have measured the depth of two uniquely identified absorption features $A$ and $B$ as $q_A$ and $q_B$ with values between 0\% and 100\%.
If the combined depth $q_A+q_B$ is larger than 100\%, than this implies that at least the larger of the two bodies (the planet) must have both chemical species in its atmosphere.

In reality, planetary spectra are far more complicated than demonstrated here, for example due to the presence of condensates in the atmosphere (clouds and haze) and spectroscopic features of other molecules.
Even in the event that the summed absorption depths of two species exceeds 100\%, the proportion of the planetary light and moon light can significantly change within the observed wavelength interval, for instance because of broad absorption by other atmospheric species.
In that case it would still be possible that one absorption trough comes from the planet and the other from a moon.
A definitive conclusion would require detailed modelling of the atmospheric properties of the planet and a possible moon.

Several techniques have been proposed that might reveal the existence of an exomoon for both transiting \cite[e.g.][]{Simon2009,Kipping2011} and directly-imaged systems \cite{Cabrera2007,Moskovitz2009,Robinson2011}.
Such techniques could provide some constraints on the interpretation of the spectra, although in most cases the detailed atmospheric properties would remain unknown.

\section{Estimate of spectral resolution} 
To estimate the spectral resolution we might expect in an observation, let us consider an Earth twin around a Sun-like star at a distance $d=10$~parsec away from the Solar System.
The flux of the star as seen from Earth is $F_* = {L_* }/(4\pi d^2)$, where $L_*$ is the star's luminosity.
In the following discussion we will assume a solar type star with solar luminosity, $L_*=3.8\cdot10^{33}$~erg/s, and temperature,~$T_*=5780$~K.

\subsection{Rate of photons}
We are interested in a specific wavelength $\lambda$ and can use Planck's law to estimate that portion of a given stellar or reflected planetary flux $F$ that is emitted in a small wavelength band~$d\lambda$ around $\lambda$:
\begin{eqnarray}
  \begin{split}
f[\lambda,d\lambda] & \approx & F\frac{\pi B_\lambda[T_*]}{\sigma T_*^4} d\lambda, \quad\quad \text{where} \\
B_\lambda[T_*] & = & \frac{2hc^2}{\lambda^5} \frac{1}{ \exp \left[ hc/(\lambda k_B T_*) \right] - 1 } \, . 
\end{split}
\end{eqnarray}
The Planck function $B_\lambda[T_*]$ is referred to as the spectral radiance, $\sigma$ is the Stefan-Boltzmann constant and $T_*$ is the effective temperature of the star. 
We can convert $f[\lambda,d\lambda]$ to a photon flux $f_\gamma$ (number of photons per area per time interval) using the relation $f_\gamma \equiv f/E = f\lambda/(hc)$, where $E = hc/\lambda$ is the energy of a photon of wavelength $\lambda$.
The rate of photons captured with an idealized telescope of diameter~$D$ and~100\% photon efficiency is then given by
\begin{eqnarray}
\dot{N} = f_\gamma \pi \, \left(\frac D2\right)^2  = F \frac{\pi^2}{4\sigma h c} \frac{\lambda \,B_\lambda[T_*]}{T_*^4}   \,D^2 d\lambda \, . \label{eq:numphoton}
\end{eqnarray}

\subsection{Spatially resolved planet}

The flux of reflected light from a spatially resolved exoplanet is a fraction of the incident stellar flux. Let us consider a planet at a distance $a$ from the star. 
If we ignore thermal radiation, the total luminosity of the planet is~$L_{\rm reflected} = L_*A\, r_p^2 /(4 a^2)$, where $A$ is the Bond albedo.
The planet is most easily observed at quadrature when its projected distance from the host star is maximized. At quadrature the planet appears as a half circle. 
The reflected light flux is usually obtained by approximating the planet's reflection properties with, for example, a Lambertian bidirectional reflectance distribution function \citep[BRDF, see e.g.][]{Seager2010}. 
Here, we use an even simpler argument which is nevertheless correct to within 10\% when compared to a Lambert sphere. 
We assume that the night side of the planet does not radiate and the star-facing side of the planet shines uniformly in each direction. 
From the perspective of Earth, we see a semicircle of the dayside and a semicircle of the nightside. 
Then, by conservation of energy, the flux of the planet as seen from Earth is simply  
\begin{eqnarray}
F_{\rm reflected} = \frac{L_{\rm reflected}}{4 \pi d^2} =\frac{L_* A r_p^2}{16\pi a^2 d^2 } \, . 
\end{eqnarray}
Note that for an Earth-like planet with a Bond albedo of~$A=0.3$ and a separation of~$a=1$~AU the contrast of the planet with respect to the star is~$F_p/F_* \sim 1.3\cdot10^{-10}$ \citep[consistent with the results of][]{Seager2010}.
We can now use Eq.~[\ref{eq:numphoton}] to calculate the rate of photons~$\dot N_{\rm reflected}$ captured with our telescope. 
If we want a signal to noise ratio of~${\rm SNR}=10$ per spectral bin in integration time~$\Delta t$, we need at least~${\rm SNR}^2=100$~photons in the wavelength band~$d\lambda$ in the Poisson noise limit, considering only noise from the planet's photons.
This condition gives us a maximum spectral resolution of 
\begin{eqnarray}
&&R_{\rm reflected}^{\rm max} =  \frac{\lambda}{d \lambda} = \frac{\lambda}{d \lambda} \; \frac{\dot{N}_{\rm reflected} \Delta t}{{\rm SNR}^2} \\
&&\quad=
\underbrace{\vphantom{\frac{\lambda^2}{T_*^4}}\frac{\pi}{64 \sigma h c}}_{\rm constants} \,
\underbrace{\vphantom{\frac{\lambda^2}{T_*^4}}\frac{A\, r_p^2 }{ a^2}}_{\rm planet}   \,
\underbrace{\vphantom{\frac{\lambda^2}{T_*^4}}\frac{L_*\lambda^2 \,B_\lambda[T_*]}{T_*^4}}_{\rm star/band} \,
\underbrace{\vphantom{\frac{\lambda^2}{T_*^4}}\Delta t \,\frac{D^2}{d^2} \; {\rm SNR}^{-2}}_{\rm telescope}  \nonumber\\
&&\quad=1683
\dep{\frac{d}{10 \rm pc}}{-2} 
\dep{\frac{D}{6.5\rm m}}{2}
\dep{\frac{\Delta t}{12\rm hrs}}{}
\dep{\frac{\rm SNR}{10}}{-2} \, . \nonumber
\end{eqnarray}
In essence, this value tells how much we can possibly learn about the planet in the most idealized observation: not enough to distinguish the two spectra presented above.
\begin{figure}[t]
\begin{center}
\resizebox{\columnwidth}{!}{\includegraphics{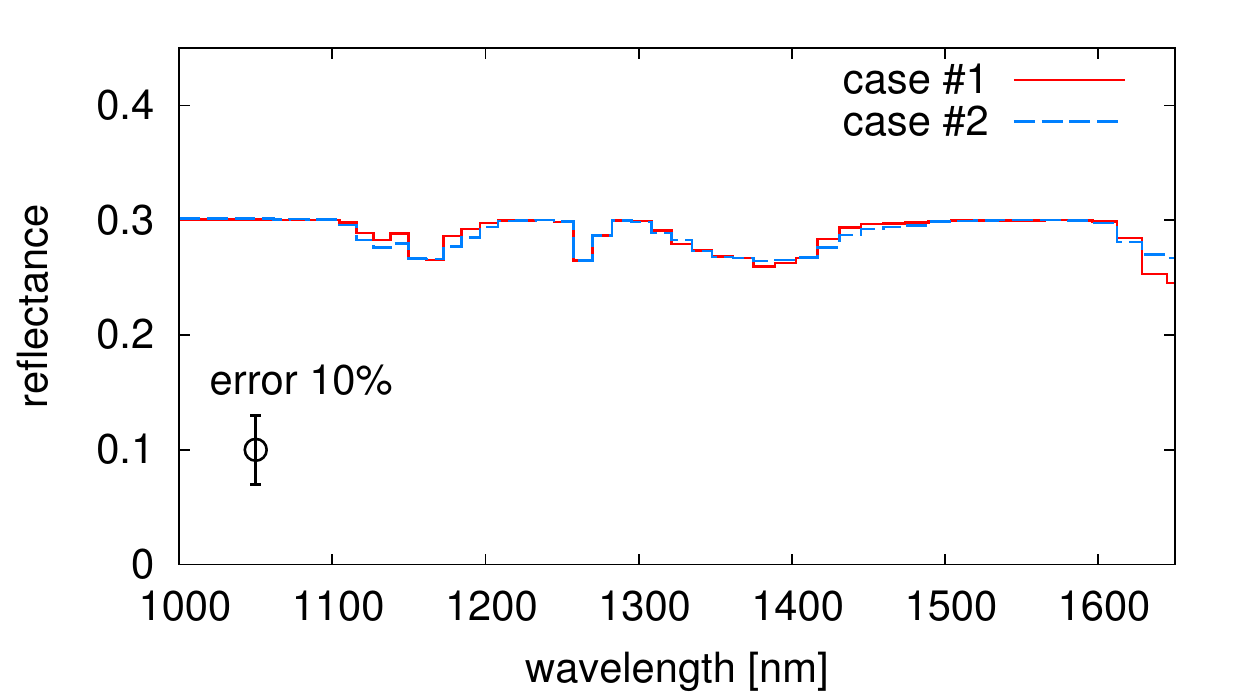}}
\end{center}
    \caption{Comparison between the low-resolution spectra ($R=100$) of case \#1 and \#2. An uncertainty bar corresponding to SNR=10, simply estimated as 10\% of the average signal, is also shown at the left bottom.\label{fig:spcomp_R100}}
\end{figure}

It is important to point out that this calculation is likely to vastly overestimate the spectral resolution for several reasons. 
For example, we set the photon efficiency of the telescope and the spectrograph to 100\%, which is clearly not realistic.
Any sort of coronagraph will reduce the throughput dramatically.
We further ignore all sources of noise except the photon noise from the planet. 
In a real observation, other sources of photon noise such as those from the host star and from exo-zodiacal light might dominate over the photon count from the planet. 
Astrophysical systematics (such as star spots) and instrumental systematics might dominate over all purely statistical noise sources.
Nevertheless, the above result gives a {\it firm upper limit} on the spectral resolution we can achieve in the best case scenario for an Earth twin.

And indeed, future space observatories such as NASA's proposed {\it Terrestrial Planet Finder} (TPF) and the European {\it Darwin} mission~\cite{kaltenegger+fridlund2005} are expected to give spectral resolutions of  only~$\lambda/d \lambda \sim 50$ \cite{Beichman2003} in the most ideal scenario.\footnote{More information about TPF and Darwin can be found at\\
\url{http://sci.esa.int/jump.cfm?oid=40843} and\\
\url{http://exep.jpl.nasa.gov/TPF-I/astrophysics.cfm}.}
Missions such as these will probably happen decades in the future.
Even then it seems highly unlikely that Earth-twin exoplanetary spectra will be achievable with significantly better SNR or spectral resolution than indicated in Fig.~\ref{fig:spcomp_R100} in the foreseeable future.

\subsection{Transiting planet}
If we are lucky enough to find a transiting Earth twin, would a transit spectrum allow a better opportunity to characterize the atmosphere than we can achieve in a high-contrast-imaging direct observation?
The technical hurdles that must be cleared to obtain a transit spectrum are much lower, as there is no need for a coronagraph to block the light from the star.
As it turns out, though, the maximum achievable spectral resolution is worse, not better, in the case of a transit spectrum.
Here, we show why.

We will consider the same Earth twin as before.
Let us further assume that the atmosphere has a scale height of~$H=7.6$~km and that~$n_H=5$ such layers will contribute to the spectrum\footnote{At the resolutions that are possible, $n_H \lesssim 5$ is a reasonable assumption that applies to both jovian \cite{hubbard_et_al2001, burrows_et_al2004} and Earth-like \cite{kaltenegger+traub2009} planets, although equation~[\ref{eq:Rmax_trans}] shows how the spectral resolution scales with $n_H$ if the reader would like to explore other values.}, resulting in an effective scale-height of~$n_HH=6.0\cdot10^{-3}r_p$.
The flux directed towards Earth during a transit and passing through the planet's atmosphere is then
\begin{eqnarray}
F_{\rm transit}  =  \frac{(r_p+n_HH)^2-r_p^2}{r_*^2}\,F_* 
  \approx  2 n_H \dep{\frac{r_p}{r_*}}{2} \dep{\frac{H}{r_p}}{} F_* , 
\end{eqnarray}
where $r_*$ is the radius of the star.
Using equation [\ref{eq:numphoton}], this corresponds to a photon rate that is much larger than in the non-transiting case
\begin{eqnarray}
\dot{N}_{\rm transit} =  \frac{8 n_HH a^2}{Ar_pr_*^2}    \dot{N}_{\rm reflected} \sim 7.3 \cdot 10^{3} \; \dot{N}_{\rm reflected} \, .
\end{eqnarray}
Initially, this looks promising for the transiting case because the signal is now given by the rate of photons passing through the atmosphere~$\dot{N}_{\rm transit}$.
However, the signal-to-noise ratio is what matters, and the noise comes from the stellar flux $\dot{N}_*$, which dominates over the flux through the planet's atmosphere by a factor of a million ($\dot{N}_* \sim  10^{6} \; \dot{N}_{\rm transit}$).
Requiring the same signal to noise ratio as above, gives us the condition ${\rm SNR}^2 = 100 = (\dot{N}_{\rm transit} \Delta t)^2 /  (\dot{N}_* \Delta t)$, or equivalently, the spectral resolution of
\begin{eqnarray}
&&\label{eq:Rmax_trans} R_{\rm transit}^{\rm max}  =   \frac\lambda{d \lambda}  = \frac\lambda{d \lambda} \; \frac{\dot{N}_{\rm transit}^2 / \dot{N}_*}{{\rm SNR}^2} \Delta t \\
&&\quad= 
\underbrace{\vphantom{\frac{\lambda^2}{T_*^4}}\frac{\pi }{4 \sigma h c}}_{\rm constants}  
\underbrace{\vphantom{\frac{\lambda^2}{T_*^4}}r_p^2 n_H^2H^2}_{\rm planet}  
\underbrace{\vphantom{\frac{\lambda^2}{T_*^4}}\frac{L_*\lambda^2 \,B_\lambda[T_*]}{r_*^4T_*^4}}_{\rm star/band} 
\underbrace{\vphantom{\frac{\lambda^2}{T_*^4}}\Delta t \,\frac{D^2}{d^2} \; {\rm SNR}^{-2}}_{\rm telescope}
\nonumber \\ 
&&\quad= 12.2
\dep{\frac{d}{10 \rm pc}}{-2} 
\dep{\frac{D}{6.5\rm m}}{2}
\dep{\frac{\Delta t}{12\rm hrs}}{}
\dep{\frac{\rm SNR}{10}}{-2} \, . \nonumber
\end{eqnarray}
Thus we have shown that the expected spectral resolution of a transiting Earth twin is extremely small. 
We might not even be able to take a spectrum of the atmosphere at all.

In particular, this result shows that taking a spectrum of a transiting Earth-like planet will be worse than that of a directly imaged analogue (putting the issue of building a coronagraph aside).
Note that taking a secondary-transit spectrum will be extremely challenging as well, since the noise is similarly inextricably dominated by the stellar photon flux.

Another factor adding to the limitations is that an Earth twin transits only once per year and for just 13 hours, setting a firm upper limit on the maximum integration time.
Also note that the transit probability for an Earth twin is $r_*/a \sim 1/200$.
Thus the closest transiting Earth twin is likely to be about $200^{1/3} \sim 6$ times farther away than the closest non-transiting Earth twin, since $200$ times the volume needs to be surveyed to find a transiting Earth twin.  
Because the distance enters the equation for the spectral resolution as $d^{-2}$, this further hurts our ability to probe the atmospheres of transiting Earth twins via their primary or secondary-transit spectra.

\section{Conclusions} 
In this article, we studied a false-positive scenario that could spoof biosignatures in spectroscopic observations of exoplanets.
We showed that a detection of two chemical species in a spectrum could be caused by light originating from two different bodies.
This is particularly important because it has been suggested that a chemical disequilibrium (which involves two or more species) could be a biomarker.
However, an observation of two species does not show that they are in fact in the atmosphere of a single object (the planet). 
An almost identical spectrum would be measured if the two species are in the atmospheres of two different bodies, one of them being on the planet and the other on the planet's moon.
Because it is impossible to resolve the moon-planet system (most likely we would not even know of its existence), the two spectra will be blended together, creating a spectrum with absorption bands of both species.

To test this scenario, we calculated synthetic spectra of an exoplanet and an exomoon, both with an atmosphere.
Using molecular oxygen and methane as the chemical species of interest, we showed that, with the spectral resolution that will be achievable with foreseeable technology, it will be impossible to tell the difference between the true-biosphere case where both species are in the same atmosphere and the false-positive case where one chemical is on the planet and the other is on the moon.
Although our specific false-positive example involves oxygen and methane, the effect is the same for any two gases that are considered a biosignature when observed together.
Our results show clearly that it is in general not possible to break the degeneracy between the two cases in a low-to-moderate resolution Earth-twin spectrum.
The only case where we can safely conclude the coexistence of two species in one planet is that the smooth continuum level is well determined and the sum of the absorption depths of two species exceeds 100\%.

We considered a large, idealized space-telescope and show that even using the most optimistic assumptions possible, the spectral resolution is unlikely to be higher than~$R\sim 1600$ for an Earth twin around a Solar-type star.
This is a fundamental physical limit just based on the photon noise.
Unless we find a planet very close to us ($d \ll 10$~pc) or develop space telescopes significantly larger than considered in this article, the only way to tweak the maximum spectral resolution is by relaxing the assumption of an Earth twin around a Solar-type star.
For example, planets that are orbiting low-mass stars and/or are somewhat larger than Earth (so-called super-Earths) have larger planet-star size ratios and could allow improved spectral resolution (see Eq.~\ref{eq:Rmax_trans}). 

Another way to avoid the exo-moon false-positive scenario altogether is to reconsider single molecule biomarkers, which do not suffer from the degeneracy presented here. 
Molecular oxygen (O$_2$) and ozone (O$_3$) have been suggested as potential single-species biosignatures.
However, either or both of them might show up in the spectrum of an abiotic planet \cite{wordsworth+pierrehumbert2014}, so they do not definitively indicate life.
Nevertheless, progress has been made in recent years on more rigorous constraints on the abiological nature of these gases~\cite{Seager2013a}.

From the perspective of the human race exploring space and searching for life on other worlds, the results of this paper are inconvenient, yet unavoidable: we will only learn the most fundamental properties of Earth twins unless we find one right in our solar neighbourhood.
It will be possible to obtain suggestive clues indicative of possible inhabitation, but ruling out alternative explanations of these clues will probably be impossible for the foreseeable future.
Since our results are based on fundamental physical laws, they are unlikely to change, even as technology advances.
The logical step forward is to widen our search for life in the universe and include objects, in the Solar System and beyond, that are less similar to Earth but more easily observable.

\begin{acknowledgments}
The work of Yuka Fujii is supported by the Grant-in-Aid No. 25887024 from JSPS. 
David S. Spiegel gratefully acknowledges support from the Association of Members of the Institute for Advanced Study.
We thank both anonymous referees for their encouraging and constructive feedback on this manuscript.
\end{acknowledgments}

{\small

 }

\end{article}

\newpage 
\ExecuteMetaData[ms_si.tex]{si}

\end{document}


\begin{article}

{\noindent\huge\titlefont Supporting Information\par}

\vspace{0.2cm}

{\noindent\large\authorfont Rein, Fujii, and Spiegel 14.xxx/pnas.xxxxxx\par}

\subsection{Model Spectra with Water Vapour}
In the main paper, we consider atmospheres that show spectroscopic signatures of only 
 O$_2$ and CH$_4$, a useful set of species to indicate the sort of chemical disequilibrium that might suggest or require life. 
In reality, however, planetary spectra are most likely far more complicated due to the effects of clouds, hazes, and the presence of other species with spectroscopic signatures. 
For instance, signatures of H$_2$O are expected to be present in Earth-like terrestrial exoplanet atmospheres.
In this Supporting Information, we present the reflection spectra of additional model atmospheres in which the planet shows additional spectroscopic signatures of H$_2$O in the atmosphere. 

The basic properties assumed for the planet and moon are the same as in Table 1 of the main paper. 
The chemical compositions of the atmospheres in our additional models are summarized in Table~S1. 
Cases \#3 and \#4 are same as cases \#1 and \#2, respectively, except that in cases \#3 and \#4 the planet contains H$_2$O in addition to O$_2$ and CH$_4$. 
For simplicity, we assume a constant (independent of altitude) H$_2$O mixing ratio of 0.15\% in case \#3 and 0.2\% in case \#4. 
This leads to a total column density of H$_2$O in our models, which is roughly equivalent to that of Earth,~$\sim$10~kg/m$^2$. 
Although we include water vapour in the atmospheres of the planet alone, it could also be present in the atmospheres of the moon or both bodies simultaneously. 

Figure~S1 compares cases \#3 and \#4 in a similar way to Figure~1 in the main paper. 
There is now only one absorption band around 1.6$\mu$m that can be uniquely attributed to CH$_4$. 
The spectra show strong absorption bands of H$_2$O around $\sim$1.15$\mu$m and $\sim$1.4$\mu$m.
These bands overlap with the CH$_4$ bands in the wavelength range shown, which could further confuse any attempt of interpretation if we have no prior knowledge of the atmosphere composition. 

The spectra of the two cases (planet vs. planet and moon), again, closely resemble each other at lower resolution. 
In fact, their resemblance is even stronger than in the cases without water vapour.
In addition, because strong absorption bands due to the planet and the moon clutter the spectra, it is hard to constrain the continuum level, making it more difficult to measure the absorption depths of oxygen and methane, which complicates one of the possible ways to break the planet/planet+moon degeneracy discussed in the main text. 

In summary, the presence of other molecules such as H$_2$O does not help us to detect a clear biosignature. Instead, additional molecules might   contribute to the confusion and could make the problem even more degenerate.

\begin{table*}[h!]
\vspace{-0.3in}
\begin{minipage}{1.5\columnwidth}
{\raggedright {\figtextfont\fignumfont Table S1.} {\fignumfont \;\;Models to Compare\hfill\\}}
\vspace{0.1cm}
\begin{tabular}{l|c|cc} 
& case \#3 & \multicolumn{2}{c}{case \#4} \\ \hline
target  & Planet & Planet & Moon  \\
composition & 15\% O$_2$ + 30ppm CH$_4$ + 0.15\% H$_2$O & 20\% O$_2$ + 0.2\% H$_2$O & 50ppm CH$_4$ \\
normalization & $r_\oplus^2$ & \multicolumn{2}{c}{$1.16r_\oplus^2$} 
\end{tabular}
\end{minipage}
\vspace{0.5cm}
\begin{center}
\includegraphics[width=\columnwidth,angle=0,clip=true]{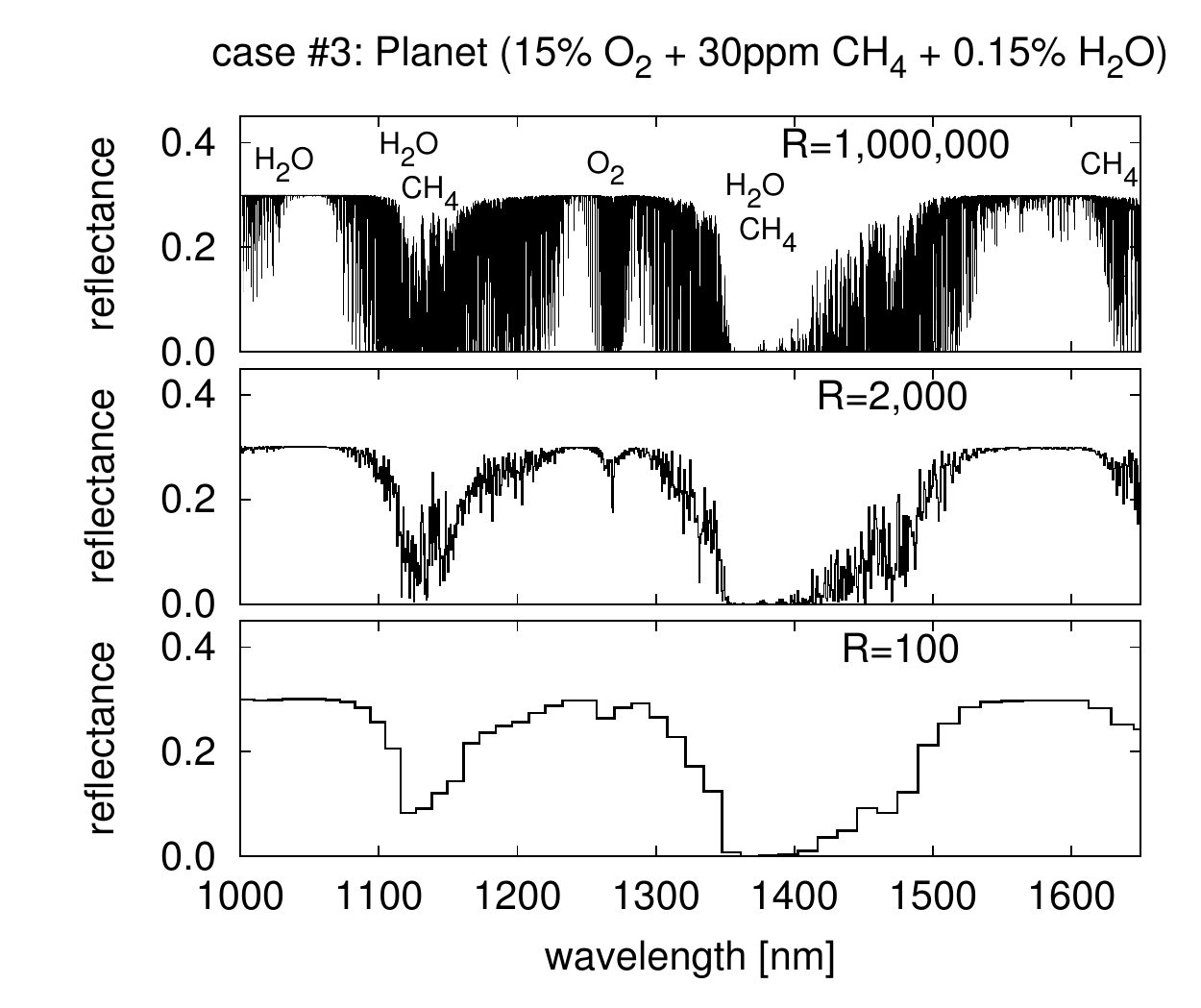}
\includegraphics[width=\columnwidth,angle=0,clip=true]{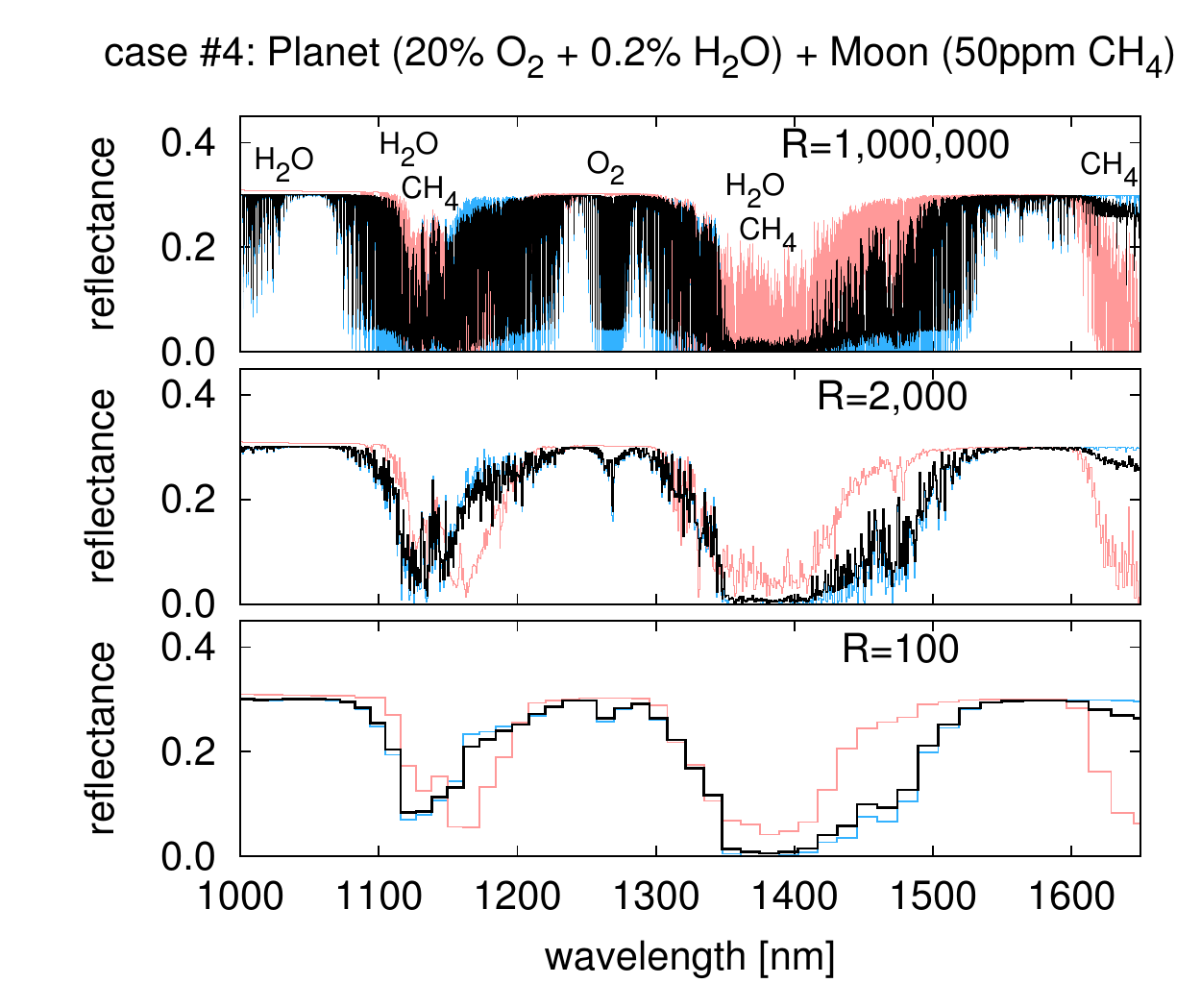}
\end{center}
\raggedright {\figtextfont\fignumfont Figure S1.} {\figtextfont Model spectra for cases \#3 and \#4 with varying resolution.
      (Left) Model spectra of a planet with 15\% O$_2$, 30ppm CH$_4$ and 0.15\% H$_2$O (case \#3).
      (Right) Black lines: Combined spectra of a planet with 20\% O$_2$ and 0.2\% H$_2$O and a moon with 50ppm CH$_4$ but without H$_2$O. 
      Blue lines: Model spectra of a planet with 20\% O$_2$ and 0.2\% H$_2$O. Red lines: Model spectra of a moon with 50ppm CH$_4$. 
      }
\end{table*}

\end{article}